\documentclass[aps,prl,reprint,groupedaddress,showpacs,floatfix]{revtex4-1}
\usepackage{graphicx}   
\usepackage{xcolor}
\usepackage{hyperref}   

\begin{document}


\title{Oscillatory elastic instabilities in an extensional viscoelastic flow}
\author{Atul Varshney, Eldad Afik, Yoav Kaplan, and Victor Steinberg}
 \affiliation{Department of Physics of Complex Systems, Weizmann Institute of Science, Rehovot, Israel 76100}

\date{\today}
\begin{abstract}
Dilute polymer solutions are known to exhibit purely elastic instabilities even when the fluid inertia is negligible. Here we report the quantitative evidence of two consecutive oscillatory elastic instabilities in an elongation flow of a dilute polymer solution as realized in a T-junction geometry with a long recirculating cavity. The main result reported here is the observation and characterization of the first transition as a forward Hopf bifurcation resulted in a uniformly oscillating state due to breaking of time translational invariance. This unexpected finding is in  contrast with previous experiments and numerical simulations performed in similar ranges of the $Wi$ and $Re$ numbers, where the forward fork-bifurcation into a steady asymmetric flow due to the broken spatial inversion symmetry was reported. We discuss the plausible discrepancy between our findings and previous studies that could be attributed to the long recirculating cavity, where the length of the recirculating cavity plays a crucial role in the breaking of time translational invariance instead of the spatial inversion.  The second transition is manifested {\it{via}} time aperiodic transverse fluctuations of the interface between the dyed and undyed fluid streams at the channel junction and advected downstream by the mean flow. Both instabilities are characterized by fluid discharge-rate and simultaneous imaging of the interface between the dyed and undyed fluid streams in the outflow channel.
\end{abstract}

\maketitle


Elastic instabilities are generic in low Reynolds number non-Newtonian fluid flows and are ascribed to the nonlinear elastic stresses generated by, e.g. polymers stretching due to the flow. Such purely elastic instabilities were observed and studied in flows with curvilinear streamlines \cite{Larson_1992, shaqfeh}. The well-studied examples of such flows are Taylor-Couette \cite{Larson_1990, groisman1}, cone-and-plate \cite{larson2} and parallel-plates \cite{larson2, groisman2, burghelea1}, Taylor-Dean \cite{shaqfeh1} and in a curvilinear channel \cite{groisman3, burghelea2, jun}. The main mechanism of instabilities in these flows is ``hoop stresses'', radially directed compressive volume forces, resulting from the azimuthal component of the elastic stresses \cite{bird, Larson_1992}. However, there are different types of flows whose instabilities are related to the existence of a stagnation point (SP), where a narrow region of highly stretched polymers extending downstream from it is formed. Such SP-related elastic instabilities were observed in axisymmetric opposed-jet device \cite{keller, keller2}, microfluidic cross-slot geometry flow \cite{Arratia_2006, Haward_2013}, T-junction \cite{mckinley}, the wake of a submerged obstacle \cite{mckinley1}, and directly related to the polymer coil-stretch transition taking place in an extensional flow at the critical Weissenberg number $Wi_c=\dot{\epsilon}\lambda=0.5$; the latter determines the degree of polymer stretching, where $\lambda$ is the longest polymer relaxation time and $\dot{\epsilon}$ is the elongation rate \cite{deGennes}.

The cross-slot elongation flow systems studied both experimentally \cite{Arratia_2006,Haward_2013} and numerically \cite{Poole_2007, Shelley_2009,graham} exhibit different bifurcations and flow properties depending on two control parameters, $Wi=\dot{\epsilon}\lambda$ and $Re=\rho\bar{\mathrm v}d/\eta$, and the details of cross-slot geometry. Here $\bar{\mathrm v}$ is the average velocity, $d$ is the width of the channel, $\rho$ and $\eta$ are the density and the viscosity of a fluid, respectively. Arratia {\it et al.} \cite{Arratia_2006} found experimentally two flow instabilities in a planar extensional flow of a dilute polymer solution at $Re<10^{-2}$ realized in a microfluidic cross-channel configuration: a forward fork-bifurcation at $Wi_{c1}\approx4.5$ resulted in a steady asymmetric flow due to the broken spatial symmetry to inversion of the inlet channels, and  a transition to aperiodic velocity fluctuations at $Wi_{c2}\approx12.5$. Poole {\it et al.} \cite{Poole_2007} reproduced numerically the transition to a steady asymmetric flow at $Re\rightarrow 0$, whereas an increase of $Re$ up to 5 delayed the onset, though not significantly. In Ref. \cite{Shelley_2009} the simulations with a steady spatially periodic background force, which simulated a four-roll mill velocity field, also revealed the steady asymmetric flow, whereas at higher $Wi$ a new state with persistent aperiodic oscillations and production and destruction of smaller-scale vortices still was not observed in experiments. In a more recent experiment \cite{Haward_2013}, a stability diagram of different bifurcations of the optimized shape cross-slot extensional rheometer (OSCER) flow in a $Wi-Re$ plane was explored using a flow birefringence for visualization of the stress field. At low $Re$ and $Wi>Wi_{c1}$, the steady asymmetric flow was observed at $El\equiv Wi/Re>1$, in accord with \cite{Arratia_2006}. However at $El<1$, a time-dependent inertio-elastic instability at $Re>Re_c\approx 18$ or larger at smaller $Wi$ (or $El$), characterized by high frequency spatio-temporal oscillations of the birefringent strand, were found \cite{Haward_2013}. It seems to be qualitatively similar to the numerical finding reported in \cite{graham}. Thus purely-elastic instabilities to a steady asymmetric flow at $Wi>3$ up to 10 and $Re$ in the range from $10^{-2}$ up to 18 was found, whereas the onset of inertio-elastic flow instabilities occurred at $Wi<10$ down to 2 and $Re$ in the range of  $\sim18$ to $200$.

The relevant paper to the results presented here is Ref. \cite{mckinley} reporting an investigation and characterization of elastic instabilities in an elongation flow in a T-junction geometry with and without a recirculating cavity at $Re<<1$ and $El>>1$. In the former case the stagnation point (SP) can freely move, whereas in the latter it is pinned at the sidewall. The unpinned SP results first in a symmetry-breaking transition to a steady asymmetric flow similar to that observed in Ref.\cite{Arratia_2006}, and then a transition to a 3D time-dependent state. However, for the pinned SP  the only transition at $Wi\simeq 3.2$ and $El\approx850$ to 3D time-dependent flow is revealed. At higher flow rates, the flow eventually becomes chaotic \cite{mckinley}.

The main result of this work is the observation and characterization of the unexpected forward Hopf transition to uniform oscillations due to the breaking of time translational invariance in the elongation flow of a dilute polymer solution flowing in a T-junction with a long recirculating cavity in a wide range of $Wi$ and  $0.1\leq Re\leq 3$. This finding is in  contrast to  previous experimental and numerical studies performed in a similar range of $Wi$ and $Re$ and reviewed above. The only difference is in the long recirculating cavity. Second instability to time aperiodic fluctuations occurs at significantly larger $Wi$. The instabilities are characterized by a measurement of the fluid discharge rate and simultaneous imaging of the interface between the dyed and undyed fluid streams in the straight outflow channel. The first transition is to transversal uniform oscillations of a flat interface between the dyed and undyed fluid streams with a frequency linearly growing with $Wi$. The second transition at higher $Wi$ manifests  as time aperiodic transverse fluctuations of the interface,  generated at the channel junction and then advected downstream by the mean flow.

\begin{figure}[t]
\begin{center}
\includegraphics[scale=0.45]{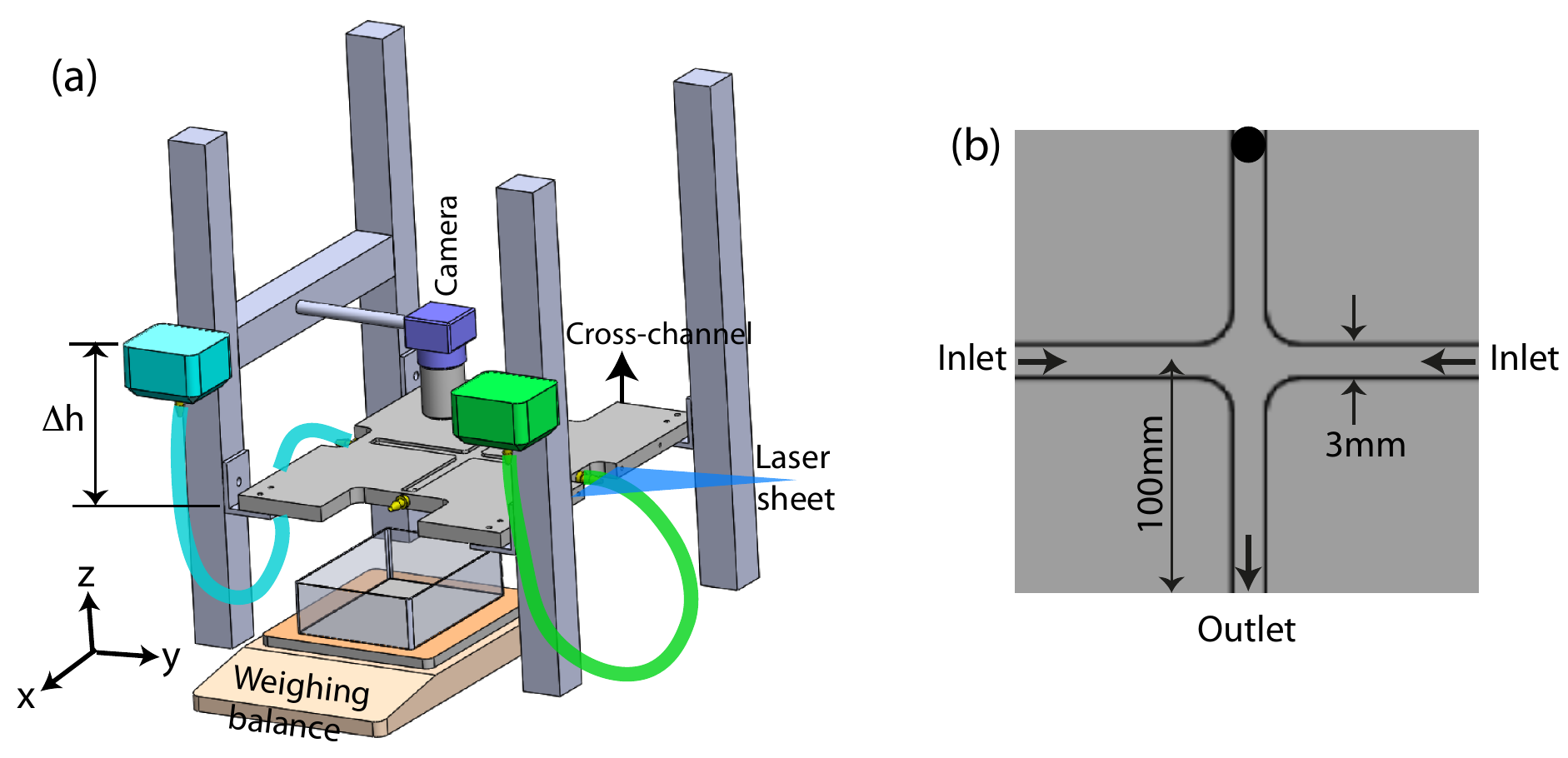}
\caption{(a) A schematic of the experimental setup (not to scale). One camera takes images of an area $19~mm\times3~mm$, $20~mm$ away from the junction, and another camera (not shown) captures an area $16~mm\times3~mm$, $74~mm$ away from the junction. (b) A schematic of the cross-slot junction.  }
\label{fig1}
\end{center}
\end{figure}

\begin{figure}[t]
\begin{center}
\includegraphics[scale=0.5]{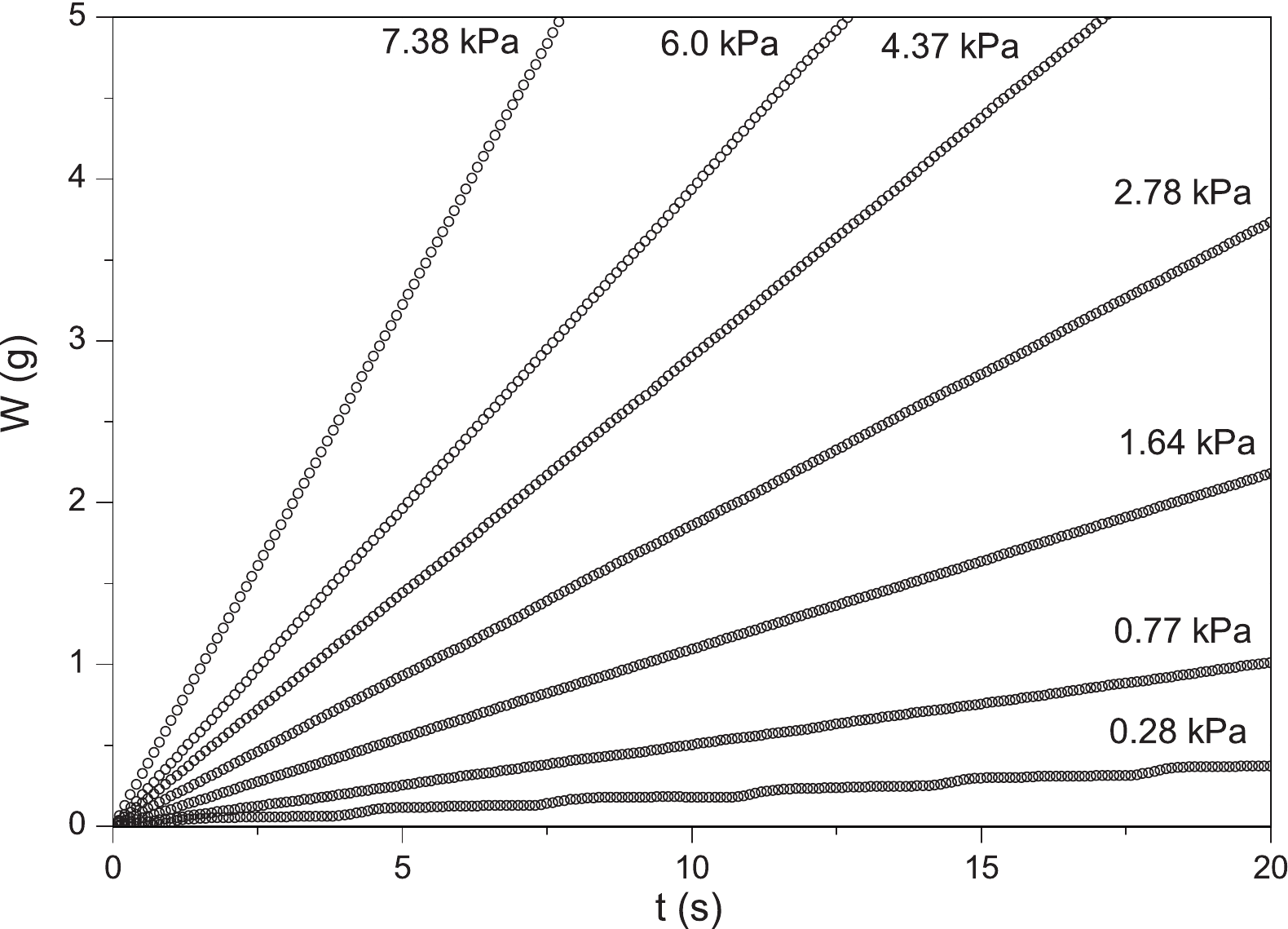}
\caption{Weight of discharged fluid ($W$) as a function of time ($t$) at different pressure difference ($\Delta P$). }
\label{fig7}
\end{center}
\end{figure}

The experimental system, shown schematically in Fig. \ref{fig1}(a) and (b), consists of a cross-channel with four arms and rounded by a radius $3~mm$ corners. Each arm is $L=100~mm$ long with a square cross-section of $d=3$ $mm$ (see Fig. \ref{fig1}(b)). Two arms served for the fluid inflow, one for the outflow, and the other outflow channel was tightly sealed  at the exit, so the SP-related elongation flow corresponds to a T-junction type with a long recirculating cavity (Fig. \ref{fig1}(b)). A dilute solution of high molecular weight polyacrylamide (PAAm, $M_w=18\times10^6$ Da; Polysciences) of concentration $c=80$ $ppm$ (and $c/c^*\simeq0.4$, where $c^*$ is the overlap concentration) was prepared in a viscous solvent of $60\%$ Sucrose and $1\%$ $NaCl$ by weight. The solvent viscosity $\eta_s$ measured in a commercial rheometer (AR-1000; TA Instruments) was found to be 0.1 $Pa\cdot s$ in the range of the shear rates $\dot{\gamma}$ from 1 to 100 $s^{-1}$ at $20^\circ$ C. The addition of the polymer to the solvent increased the solution viscosity $\eta$ by about 30$\%$, which in its turn revealed a shear thinning by about $30\%$ in the range of $\dot{\gamma}$ from 3 to 100 $s^{-1}$ at $20^\circ$ C. The stress-relaxation method \cite{Liu_2007} was employed to measure $\lambda=10\pm0.5~s$. The polymer solution premixed with 10 $ppm$ of a Fluorescein dye was injected into one of the two inlet channels. Both the dyed and undyed polymer solutions flow was gravity-driven, and its rate was varied by changing the hydrostatic pressure $\Delta P=\Delta h \rho g$ {\it{via}} $\Delta h$, as shown in Fig. \ref{fig1}(a). Here $\rho=1286$ $Kg/m^3$ is the density of the solution, and $g$ is the gravitational acceleration. The fluid exiting  from the outflow channel was weighed instantaneously $W(t)$ as a function of time $t$ (see Fig. \ref{fig7}) by a balance PC-interfaced with a sampling rate $10~Hz$ and a resolution of $1~mg$ (BPS-1000-C2; MRC). The average fluid discharge rate $\bar{Q}$ was estimated as $\bar{Q}={\overline{\Delta W}/\Delta t}$, where an over bar denotes time averaging. The Weissenberg number is further defined as $Wi=\lambda \dot \gamma$, where $\dot \gamma\simeq 4\bar{Q}/(\rho d^3)$ is the estimated shear-rate in a 3D square channel flow, $\bar{\mathrm v}=\bar{Q}/(\rho d^2)$ is the average flow speed, and the coefficient $\sim2$ is the ratio of ${\mathrm v}_{max}/\bar{\mathrm v}\simeq2$ calculated from an approximate solution of the Stokes flow in a square cross-section channel \cite{Brenner}.

\begin{figure}[t]
\begin{center}
\includegraphics[scale=0.7]{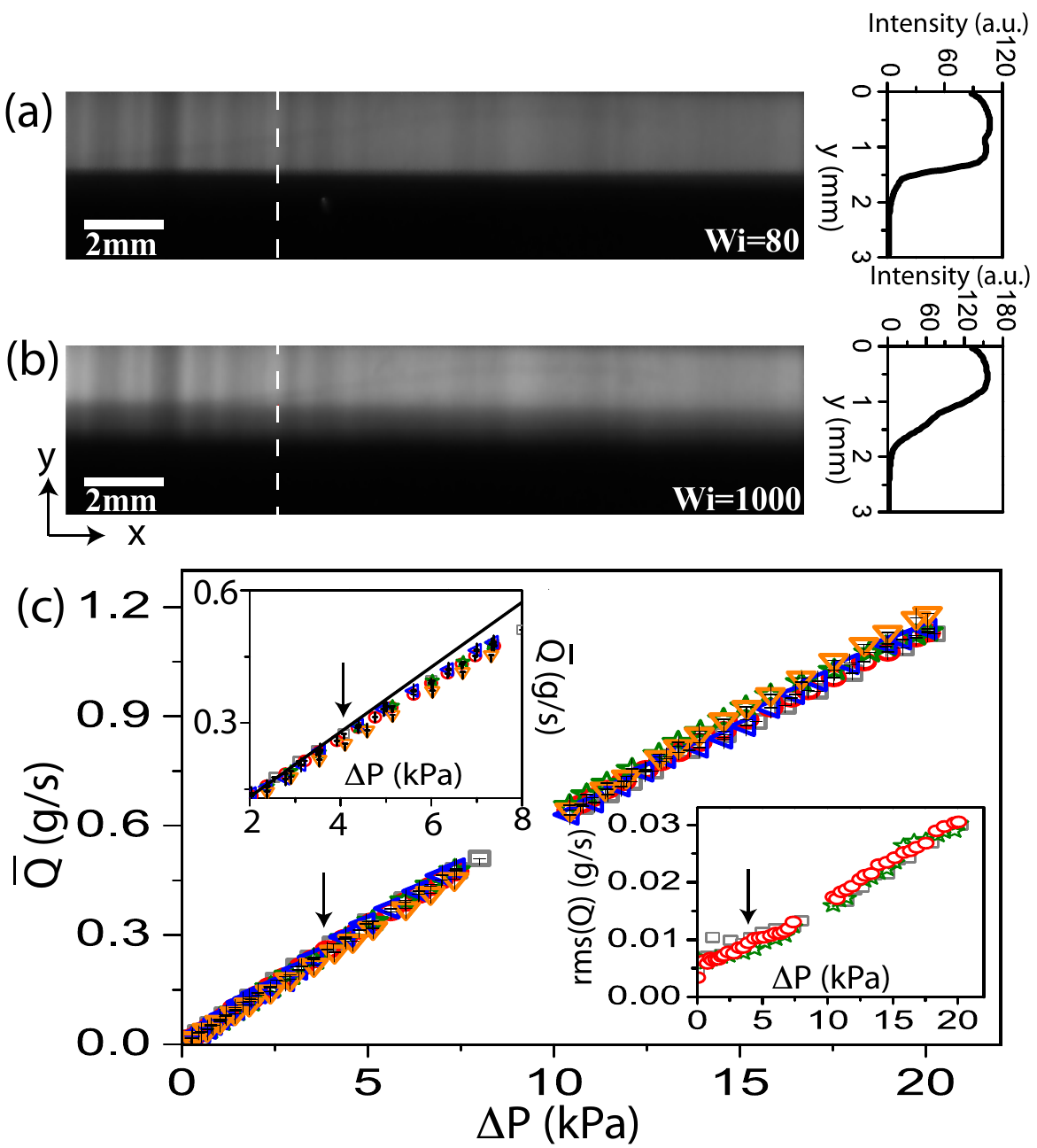}
\caption{(a) and (b) Snapshots of the flow and the corresponding intensity profile at $x=25.5~mm$ from the junction (marked by dashed line). (c) $\bar{Q}$ versus $\Delta P$. Top inset: a magnified view of the data close to the onset of the first instability ($\Delta P\sim 4~ kPa$) shown by the deviation of the data from its initial slope indicated by a solid line at $\Delta P<4~ kPa$. Bottom inset: rms $Q$ versus $\Delta P$. Different symbols are for different runs. The arrows on the main plot and both insets indicates the location of the first transition.}
\label{fig2}
\end{center}
\end{figure}

Two regions of the outflow channel, close to the  junction and near the exit, are illuminated uniformly by a laser sheet of $\sim 100$ $\mu m$ thickness generated by a combination of two cylindrical and one plano-convex lenses and an Ar-ion laser beam at 488 $nm$ wavelength. The dynamics of the interface between the dyed and undyed flow streams in the outflow channel close to the  junction as well as near the exit is recorded by two CCD cameras (EC1380, Prosilica and GRAS 14S5M-C, Point Grey) with bandpass filters XF1074 (the camera near the exit is not shown in Fig. \ref{fig1}(a)). An optical chopper at a frequency 23 $Hz$ is used to reduce heating of the system due to the laser illumination, and both cameras are synchronized at the chopper frequency. Figure \ref{fig2}(a) and (b) shows snapshots of the flow captured near the junction at $Wi=80$, $Re=0.176$ and $Wi=1000$, $Re=2.2$ respectively.  The coordinates of the interface are identified by an estimator for the extremum in the derivative of the spatial variation in a light intensity with respect to y-axis, as shown next to each snapshot in Fig. \ref{fig2}(a) and (b).
Two regions of the outflow channel, close to the  junction and near the exit, were illuminated uniformly by a laser sheet of $\sim 100$ $\mu m$ thickness generated by a combination of two cylindrical and one plano-convex lenses and an Ar-ion laser beam at 488 $nm$ wavelength. The dynamics of the interface between the dyed and undyed flow streams in the outflow channel close to the  junction as well as near the exit was recorded by two CCD cameras (EC1380, Prosilica and GRAS 14S5M-C, Point Grey) with bandpass filters XF1074 (the camera near the exit is not shown in Fig. \ref{fig1}(a)). An optical chopper was used at a frequency 23 $Hz$ to reduce heating of the system due to the laser illumination, and both cameras were synchronized at the chopper frequency. Figure \ref{fig2}(a) and (b) shows snapshots of the flow captured near the junction at $Wi=80$, $Re=0.176$ and $Wi=1000$, $Re=2.2$ respectively.  The coordinates of the interface are identified by an estimator for the extremum in the derivative of the spatial variation in the light intensity with respect to the y-axis, as shown next to each snapshot in Fig. \ref{fig2}(a) and (b).

\section*{Results and discussion}
Figure \ref{fig7} shows the variation of weight of the discharged fluid ($W$) as a function of time ($t$) for several values of the pressure difference ($\Delta P$). At low value of $\Delta P$ ($\sim0.28~kPa$), small amplitude oscillations resulting from the dripping of fluid drops can be noticed. However, at higher values of $\Delta P$ ($>0.7~kPa$),  the fluid fall in a continuous stream,  thereby the dependency W(t) becomes smooth and no oscillations are observed.    
Figure \ref{fig2}(c) shows the dependence of $\bar{Q}$ on $\Delta P$ for several data sets. At low pressure, the dependence is linear as expected for laminar flows. At $\Delta P\sim4$ $kPa$, the dependence deviates slightly from the initial slope indicating the onset of the first instability (see the top inset of Fig. \ref{fig2}(c) where the arrow indicates approximately the onset of the first transition). The second instability appears  above $\Delta P\sim 10$ $kPa$ that is manifested by a more significant change in the slope. The gap in the data points at $\Delta P\simeq 7.5-10$ $kPa$ is due to the apparatus design. The transitions are pronounced in the rms of $Q$, which grows with a greater rate with $\Delta P$ above the second instability (see the bottom inset in Fig. \ref{fig2}(c) where the arrow indicates same as before). The error bars in $\bar{Q}$ represent the $95\%$ confidence interval.

\begin{figure}[t!]
\begin{center}
\includegraphics[scale=0.45]{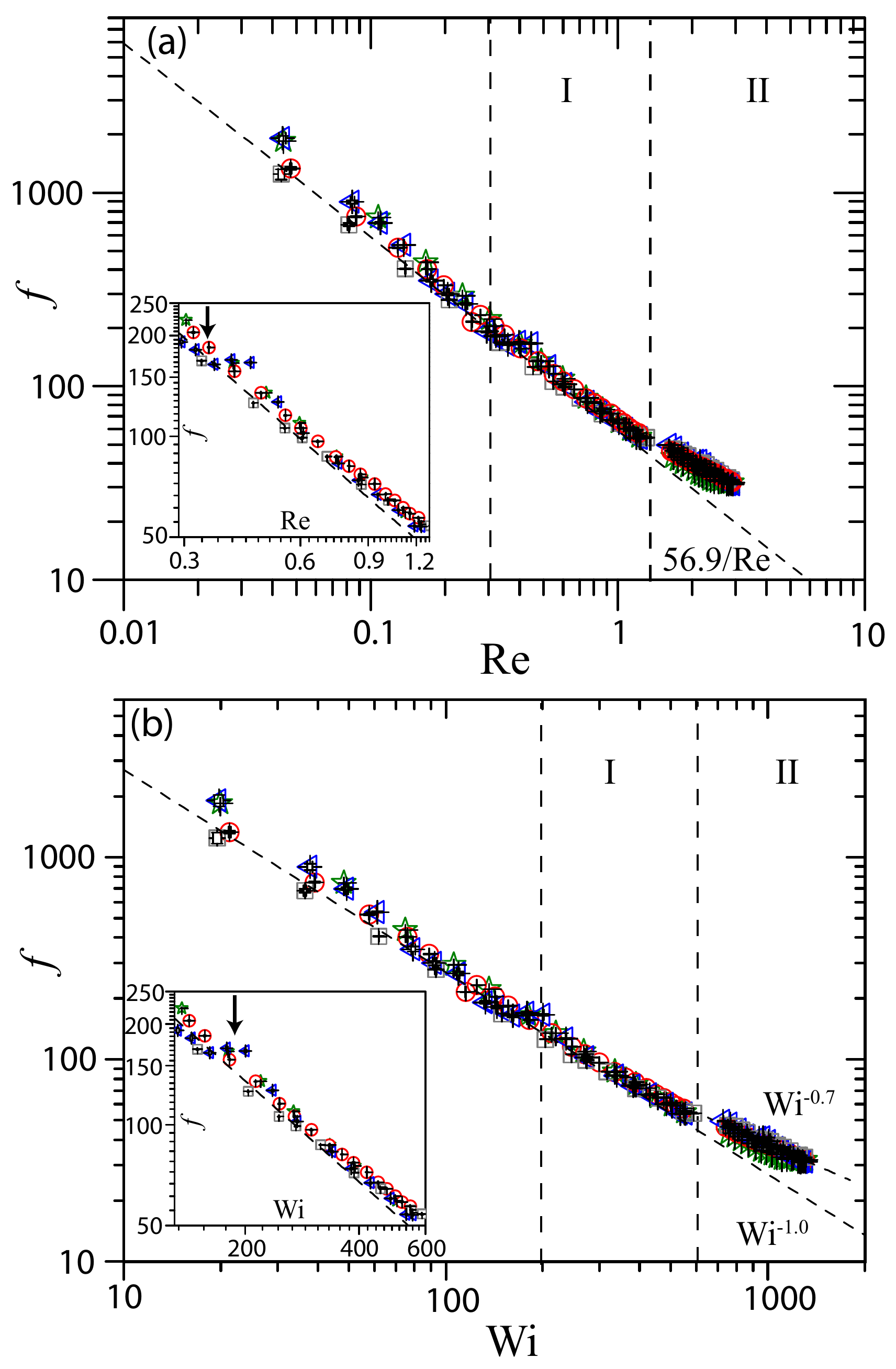}
\caption{(a) Friction factor $f$ versus $Re$. The dashed line is a theoretical dependence $56.9/Re$ for a laminar flow of a Newtonian fluid in a square channel. The inset: a magnified view of the data close to the onset of the first instability ($Re_{c1}\sim 0.3$). (b)  $f$ versus $Wi$. In the laminar regime $f$ varies as $Wi^{-1.0\pm 0.02}$, and above the second instability $f\sim Wi^{-0.7\pm0.01}$. The dashed lines are fits to the data with different symbols for different runs. The inset: a magnified view of the data close to the onset of the first instability ($Wi_{c1}\sim 200$). Regions of the first oscillatory instability $\mathrm{I}$ and the second instability of aperiodic oscillations $\mathrm{II}$ are indicated on the main plots by vertical dash lines, respectively. }
\label{fig3}
\end{center}
\end{figure}

In order to highlight more clearly the transitions, the data shown in Fig. \ref{fig2}(c) are presented on a high resolution plot in Fig. \ref{fig3}(a,b) via the dimensionless friction factor defined as $f=(\Delta P_c\times d)/(\rho \bar{{\mathrm v}}^2L/2)$. Here the pressure drop along both a single inflow and outflow channels $\Delta P_c=\Delta P-\Delta P_{pipe}$, $\Delta P_{pipe}$ is the pressure drop along the feeding pipe connecting the fluid container to the inlets of the cross-channel, having an inner diameter $D_p=4.7~mm$ and length $L_p=2.15$ $m$. $\Delta P_{pipe}$ is calculated using the expression for the friction coefficient for a pipe laminar flow of Newtonian fluid\cite{Moody_1944} $f_{pipe}=64/Re_{pipe}$, where $Re_{pipe}=D_p\bar{v}_{pipe}\rho/\eta$, $\bar{v}_{pipe}=2\bar{Q}/(\rho\pi D_p^2)$, and as a result one gets $\Delta P_{pipe}=(64\bar{Q}\eta L_p)/(\rho\pi D_p^4 )$. Since in a straight pipe of sufficiently large diameter $D_p$ polymers remain mostly unstretched, the same formula for $f$ can be used for a laminar flow of polymer solutions.

To further proceed with the calculations of $f$ along solely the outflow channel, one should take into account that both the inlet and outlet channels have the same cross-section and length but the flow discharge through the inlet channel is half that of the outlet. Then the correct expression for the friction factor of solely the outflow channel is $f=4\Delta P_c \rho d^5/(3L \bar{Q}^2)$ presented as a function of $Re$ in Fig. \ref{fig3}(a). In the laminar regime at $Re<1$, the theoretical dependence of $f$ on $Re$ for a channel laminar flow of Newtonian fluids with a square cross-section\cite{Jones_1976} is $f=56.9/Re$  and agrees well with the data shown in Fig. \ref{fig3}(a) without any fitting parameters. The deviation from this dependence is visible at $Re\geq 0.5$. Since in the vicinity of the first bifurcation $Re\simeq 0.3$, the inertial effect is negligible and the control parameter in the problem appears to be $Wi$. The same data in $f$ versus $Wi$ coordinates are presented in Fig. \ref{fig3}(b). Then the first transition is identified by a weak deviation from about the linear decay at $Wi\approx 200$, and the second one at $Wi\approx 700$ towards $f\sim Wi^{-0.7\pm0.01}$ dependence. In the entire range of $Re$ and $Wi$ explored in the experiment one gets $El\gg 1$.

\begin{figure}[t]
\begin{center}
\includegraphics[scale=0.7]{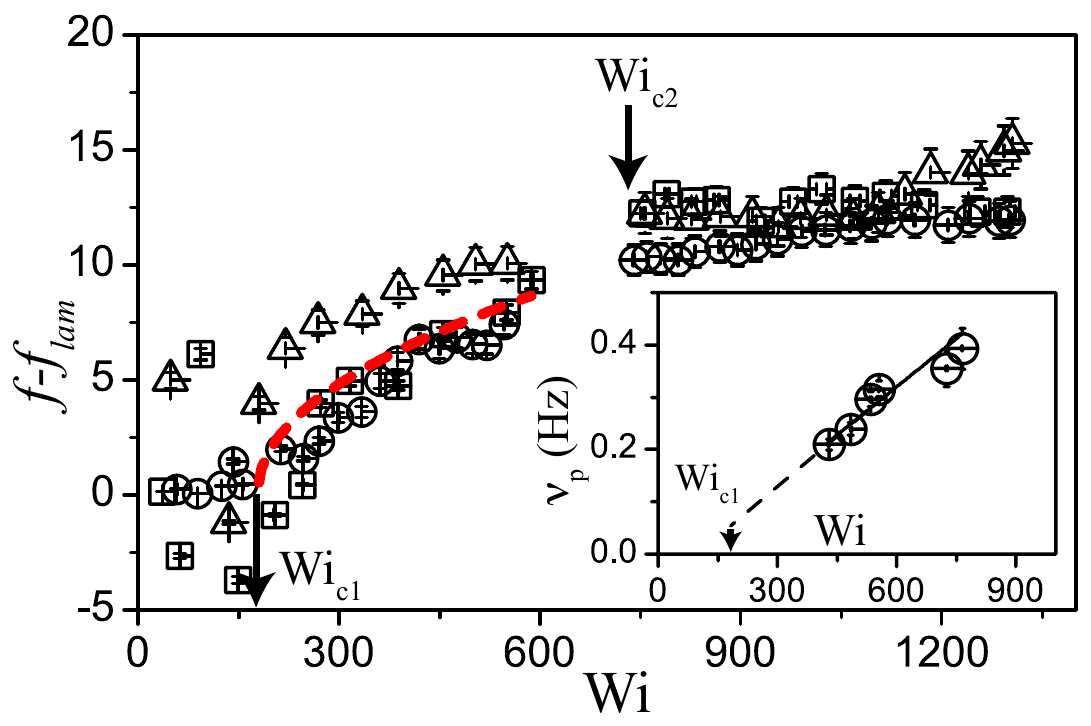}
\caption{$f$ reduced by $f_{lam}$ versus $Wi$ with different symbols for different runs. The dashed line is a squared-root fit to the data. Arrows indicate the onset of both the first and second instabilities. Inset: $\nu_p$ versus $Wi$ above $Wi_{c1}$. The solid line is a linear fit to the data.}
\label{fig4}
\end{center}
\end{figure}

To better  elucidate the nature of the transitions we subtract the laminar flow contribution $f_{lam}$ for each data set from the friction factor $f$, i.e., $f-f_{lam}$, and plot it versus $Wi$ in Fig. \ref{fig4}. Since this presentation further amplifies the plot resolution, the random scatter in the data as well as the systematic errors between the different data sets increase. In spite of this, one finds the first transition with characteristics of a forward bifurcation indicated  by the squared-root fit for all data sets, which yields critical Weissenberg numbers $Wi_{c1}\simeq180\pm20$ for the onset of the first instability. Due to the gap in the data, the onset of the second instability is determined with a significant uncertainty in the range of $Wi_{c2}\sim 650-750$. We would like to point out here a discrepancy with the results of previous experiments in the values of $Wi_{c1}$ and $Wi_{c2}$, which are larger by about two orders of magnitude. In the current presentation $Wi$ is defined via the shear rate $\dot{\gamma}$, in contrast to previous reports, where either the measured or the estimated elongation rate $\dot{\epsilon}$ was used. The ratio between $\dot{\gamma}$ and $\dot{\epsilon}$ can differ up to 10 times and more, as pointed out in Ref. \cite{Haward_2013}, this increases the values of $Wi$ by the same factor. Another reason for the large values of $Wi_{c1}$ and $Wi_{c2}$ may be the macroscopic size of the channel in our experiment compared with microfluidic channels in the previous ones.

\begin{figure}[t]
\begin{center}
\includegraphics[scale=0.75]{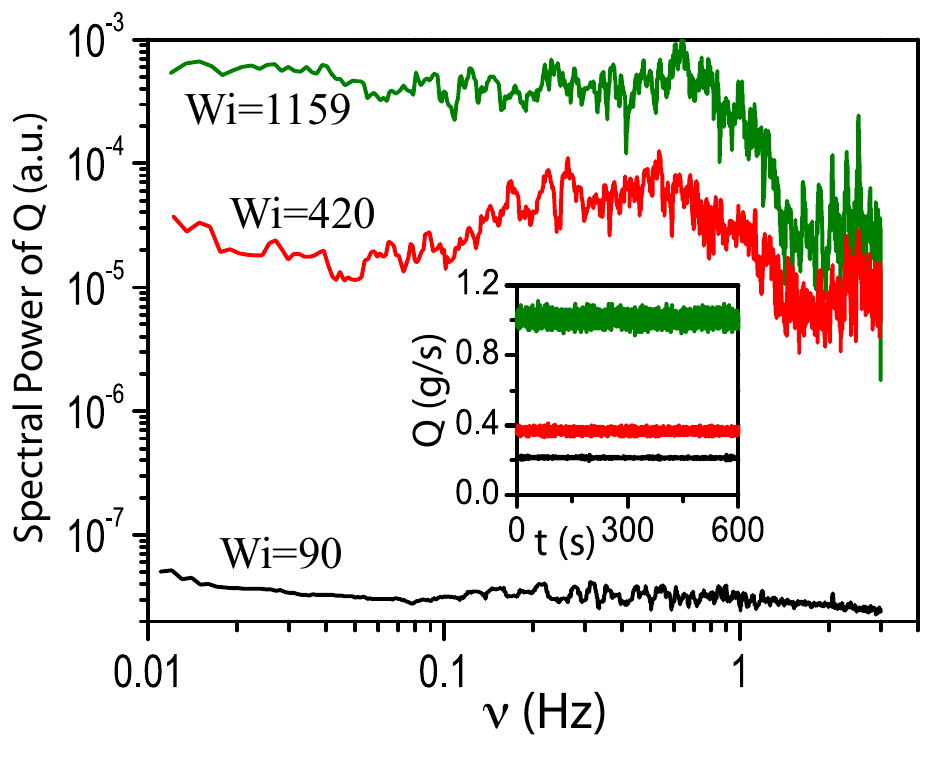}
\caption{Power spectra of $Q$ for various $Wi$. Inset: $Q$ versus $t$ at the same $Wi$.}
\label{fig5}
\end{center}
\end{figure}

To characterize the nature of the instabilities above the first and second transitions we study the frequency power spectra of the discharge rate $Q(t)$. The $Q$ time series at different $Wi$ are presented in the inset in Fig. \ref{fig5}, from which the respective power spectra are computed. The main plot in Fig. \ref{fig5} shows typical frequency power spectra of $Q$ in the regimes of the laminar flow, above the first and second instabilities.  The spectral power for $Wi=420$ increases about three orders of magnitude at low frequencies and shows a broad peaks around $0.3~Hz$ with a consequent frequency cut-off, probably due to a mechanical inertia of the balance measurements, at frequencies above about $1~Hz$ in contrast to the laminar flow at $Wi=90$, whose spectrum is rather flat. At $Wi=1159$, the spectrum turns flat again at a spectral power an order of magnitude larger than that at $Wi=420$ and with a frequency cut-off similar to the one observed at $Wi=420$. 

\begin{figure}[t]
\begin{center}
\includegraphics[scale=0.52]{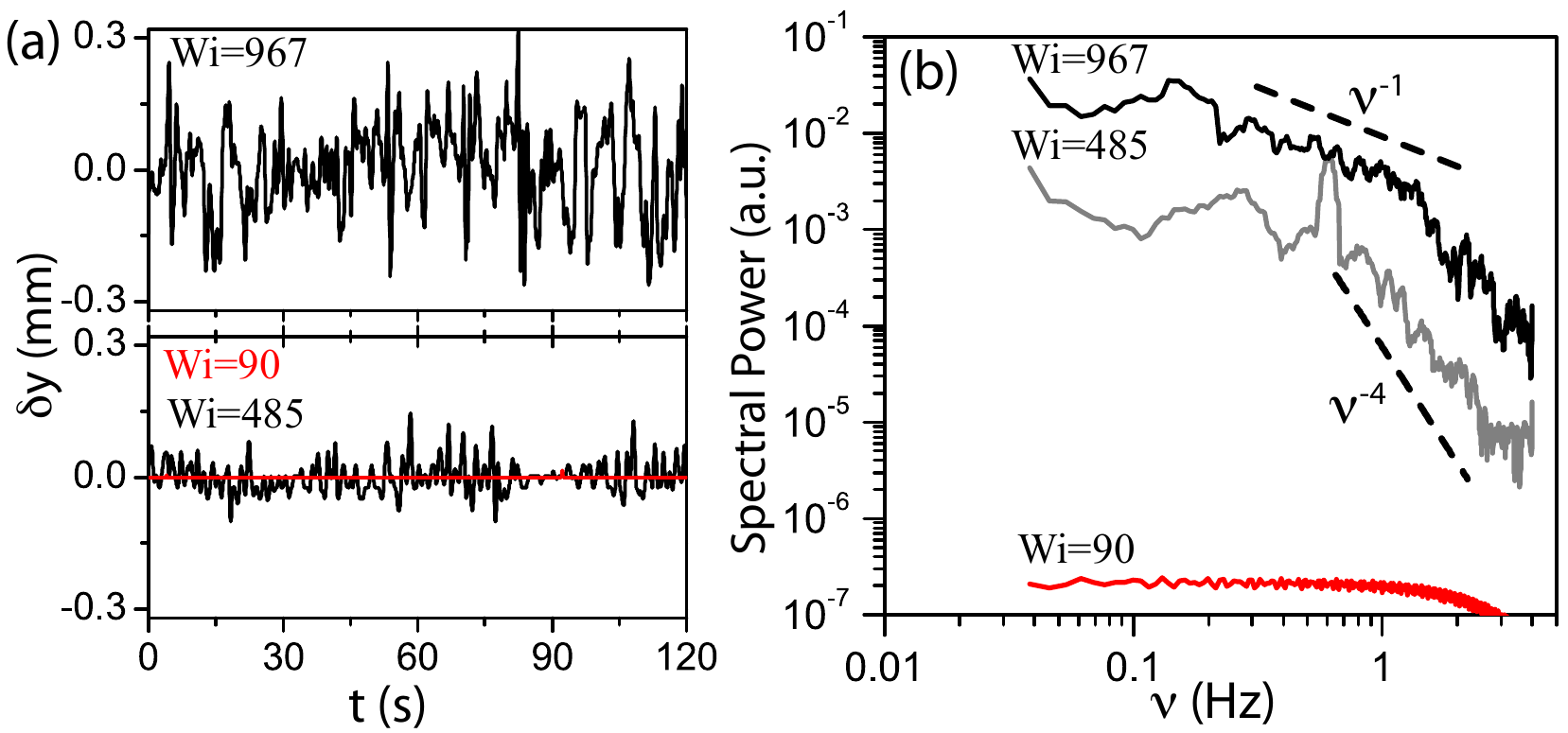}
\caption{(a) $\delta y$ versus $t$ at $Wi=90$ in a laminar regime (red line), at $Wi=485>Wi_{c1}$, and at $Wi=967>Wi_{c2}$. (b) Power spectra of $\delta y$ for the same $Wi$. The dashed lines show: $1/\nu$ for aperiodic fluctuations at $Wi>Wi_{c2}$, and the cut-off scaling  $\nu^{-4}$ for both spectra at $Wi>Wi_{c1}$ and $Wi>Wi_{c2}$.}
\label{fig6}
\end{center}
\end{figure}

Finally, we characterize the instabilities by measuring a motion of the interface between the dyed and undyed fluid streams of the polymer solution.  The interface location is identified at a given point along the outflow channel by an estimator for the extremum of the intensity spatial derivative across the channel width. In the regime of the laminar flow, the interface is sharp (see Fig. \ref {fig2}(a)) and remains steady for a long time as shown by temporal interface fluctuations from its mean value $\delta y(t)$  and in the corresponding power spectrum for $Wi=90$ in Fig. \ref{fig6}(a,b). 
Above the first bifurcation the interface commences to oscillate uniformly. It is reflected in a strong growth of $\delta y$ (see Fig. \ref{fig6}(a) at $Wi=485$) and characterized by a broad main peak  with a frequency $\nu_p\sim0.3~Hz$, its sharp second harmonic peak at $\nu_p\sim0.6~Hz$ in the power spectrum at $Wi=485$ (Fig. \ref{fig6}(b)),  and  about four orders of magnitude growth in the spectral power at low frequencies. The characteristic interface oscillation frequency $\nu_p$ of the peak in the power spectra as a function of $Wi$ is presented in the inset in Fig. \ref{fig4}. It grows linearly with $Wi$ and its extrapolation to zero gives $Wi_c$ rather close to the value found from the fit of $f-f_{lam}$ versus $Wi$ shown on the main plot in Fig. \ref{fig4}. Both properties characterize the first transition as the forward Hopf bifurcation \cite{cross}. Above the second instability, the interface motion becomes aperiodic in time with sporadic transverse perturbations of the interface generated at the T-junction and then advected downstream. This leads to another strong increase of $\delta y$ in time series data shown in Fig. \ref{fig6}(a) at $Wi=967$ and results in a power-law decay of about $\nu^{-1}$ in the power spectrum (Fig. \ref{fig6}(b)). At low frequencies, the spectrum amplitudes grow  larger than one order of magnitude, and at high frequencies both spectra at $Wi=485$ and $967$ have an algebraic cut-off of about $\nu^{-4}$.

 The main unexpected result reported in the paper is the forward Hopf bifurcation at $Wi_{c1}\simeq 180$ resulting in a uniformly oscillating state due to breaking of the time translational invariance \cite{cross}, in contrast to the forward fork-bifurcation into a steady asymmetric flow due to the broken spatial inversion symmetry  found in previous studies\cite{Arratia_2006, Haward_2013, Poole_2007, Shelley_2009, graham, mckinley}.

 There are two possible factors which may be responsible for the unexpected observation of the first transition as a Hopf bifurcation: (i) the value of $Re$, which is about 0.3 at the transition compared with about $10^{-2}$ realized in a microfluidic cross-channel and T-shaped configurations \cite{Arratia_2006,Haward_2013,mckinley}; (ii) the long recirculating  cavity in contrast to the short one used in \cite{mckinley}, where the steady fork bifurcation to the asymmetric steady flow is observed. Regarding the first factor, as shown by the numerical simulations\cite{Poole_2007} the higher values of $Re$ up to 5 and by the experiment\cite{Haward_2013} the values of $Re$ till 18 do not change the nature of the transition to a steady asymmetric flow but just insignificantly delays the onset. On the other hand, a long recirculating cavity, compared to the one used in the experiment reported in Ref. \cite{mckinley}, has not been explored previously, either in experiment or numerical simulations. Thus the length of the recirculating cavity plays a crucial role in the breaking of time translational invariance instead of spatial inversion. Intuitively, it looks more energetically demanding to switch to a steady asymmetric flow in a long recirculating cavity rather than to uniform oscillations. We believe that our main finding, the characterization of the first transition as a Hopf bifurcation,  in contrast to  previously reported  experimental and numerical studies, as well as the role of the recirculating cavity length on the flow behavior will trigger interest to perform numerical simulations of the flow configuration used in this experiment.

This work was partially supported by grants from Israel Science Foundation and Volkswagen Foundation via the Lower Saxony Ministry of Science and Culture Cooperation (Germany). A. V. acknowledges the support of the VATAT program for a postdoctoral fellowship.
\bibliography{ei} 

\end{document}